\documentstyle[aps,prl]{revtex}
\begin{document}
\title{Unconventional Ferromagnetic Transition in La$_{1-x}$Ca$_x$MnO$_3$}
\author{J. W. Lynn,$^1$ R. W. Erwin, J. A. Borchers, Q. Huang,$^1$ and A. Santoro}
\address{Reactor Radiation Division, National Institute of \\
Standards and Technology, Gaithersburg, Maryland 20899}
\author{J-L. Peng and Z. Y. Li,}
\address{Center for Superconductivity Research, University of Maryland,$^1$ College\\
Park,\\
MD 20742}
\date{\today  }
\maketitle

\begin{abstract}
Neutron scattering has been used to study the magnetic correlations and long
wavelength spin dynamics of La$_{1-x}$Ca$_x$MnO$_3$ in the ferromagnetic
regime ($0\leq x<\frac 12$). For $x=\frac 13$ ($T_C=250K)$ where the
magnetoresistance effects are largest the system behaves as an ideal
isotropic ferromagnet at low T, with a gapless ($<0.04meV$) dispersion
relation $E=Dq^2$ and $D_{T=0}\approx 170$ meV-\AA $^2$. However, an
anomalous strongly-field-dependent diffusive component develops above $\sim
200K$ and dominates the fluctuation spectrum as $T\rightarrow T_C$. This
component is not present at lower $x$.

\end{abstract}

\twocolumn
The magnetic properties of the doped LaMnO$_3$ class of materials have been
under very active investigation recently because of the dramatic increase in
the conductivity when the spins order ferromagnetically, either by lowering
the temperature or applying a magnetic field. This huge increase in the
carrier mobility originates from a metal-insulator transition that is
closely associated with the magnetic ordering. The largest effect is found
for the calcium-doped system, and therefore initially we have concentrated
on bulk La$_{1-x}$Ca$_x$MnO$_3$ materials.\cite{GeneralRef} We find that the
system is ferromagnetic for x$<$0.5, including the undoped sample, in
agreement with recent results for this material.\cite{Cheong1} For the x=$%
\frac 13$ doping where the magnetoresistance anomalies are largest\cite
{Jin94} we find that at low T the magnetic system behaves as an {\em ideal }%
isotropic ferromagnet. At elevated temperatures, however, a quasielastic
component to the fluctuation spectrum develops for the $x=\frac 13$ doping,
and becomes the dominant spectral weight as $T\rightarrow T_C$. This
behavior is in stark contrast both to the conventional behavior observed for
isotropic ferromagnets and for the Ca-doped materials away from $x=\frac 13$%
. The width of this scattering is proportional to $q^2$, indicating that it
represents spin diffusion. The correlation length, on the other hand, is
anomalously small ($\sim 10\AA $) and only weakly temperature dependent.
These results suggest that this quasielastic component is associated with
the electrons becoming localized on the Mn$^{3+}$/Mn$^{4+}$ lattice, and it
is this spin diffusion that drives the ferromagnetic phase transition rather
than the thermal population of conventional spin waves.

The diffraction, small angle neutron scattering, and inelastic experiments
were all carried out at the NIST research reactor. Because the long
wavelength spin dynamics turn out to be approximately isotropic, inelastic
measurements on polycrystalline samples may be made in the forward
scattering direction (i.e. around the (000) reciprocal lattice point)
without loss in generality.\cite{amorphous} All the inelastic measurements
reported here were taken on the BT-9 triple-axis spectrometer, with
pyrolytic graphite monochromator, analyzer, and filter. The incident energy
was chosen to be 13.7 meV, and horizontal collimations of 12$^{\prime }$-11$%
^{\prime }$-12$^{\prime }$-16$^{\prime }$ full width at half maximum (FWHM)
were used. The use of polycrystalline samples has the distinct advantage
that powder diffraction/profile refinements, which establish the oxygen
content and provide detailed crystallographic parameters, may be performed
on the same samples. Presently we have carried out measurements for Ca-doped
polycrystalline samples of x = 0, 0.15, 0.175, 0.33, 0.44, and 0.5 made by
the usual solid state reaction technique. Our crystallographic studies
reveal that there can be orthorhombic, rhombohedral, and monoclinic phases
in this range of doping both as a function of temperature and as a function
of doping, and these three phases represent subtle structural distortions of
the basic perovskite structure. We also observe anomalies in the lattice
parameters and structures that are associated with the magnetic ordering, as
has recently been reported.\cite{Distortion} However, we believe that these
distortions do not by themselves strongly influence the basic spin dynamics
and we will discuss the crystallography in detail elsewhere.\cite{Huang}

For the x=$\frac 13$ ($T_c$ = 250 K) material that is of primary interest
here the crystal structure is orthorhombic (Pnma) over the entire range of
temperature and field we have explored, while for temperatures well below $%
T_c$ the magnetic system behaves as an ideal isotropic ferromagnet. Thus in
the long wavelength (hydrodynamic) regime the magnetic excitations are
expected to be conventional spin waves with a dispersion relation $E=\Delta
+D(T)q^2$, where $\Delta $ represents the spin wave energy gap and the spin
stiffness coefficient $D(T)$ is directly related to the exchange
interactions.\cite{HandH} Fig. \ref{fig1}(a) shows  
a typical magnetic inelastic spectrum collected at $200K$ and wave vector $%
q=0.07\AA ^{-1}$. A flat background of 9 counts plus an elastic incoherent
nuclear peak of 647 counts, measured at 14K, have been subtracted from these
data. We see that the spectrum is dominated by spin waves observed in energy
gain (E%
\mbox{$<$}
0) and energy loss (E%
\mbox{$>$}
0). Similar data were obtained at a series of wave vectors and temperatures,
and Fig. 1(b) shows the measured spin wave dispersion relation at two
different temperatures. We see that the dispersion relation is indeed
quadratic in $q$. At low temperature the widths of the spin waves are solely
instrumental in origin, and this demonstrates that the dispersion relation
is indeed isotropic in $\overrightarrow{q}$; any significant anisotropy
would reveal itself as an apparent intrinsic linewidth in these powder
measurements.\cite{anisotropy} This isotropy in the dispersion relation
indicates that the net exchange interactions do not depend on the direction
in the crystal to a good approximation. We also note that any spin wave gap $%
\Delta $ is too small (%
\mbox{$<$}
0.04 meV) to be determined in these measurements, which is consistent with
high resolution measurements on a Sr-doped single crystal.\cite{Doloc} The
energy gap $\Delta $ may be simply interpreted as the cost in energy to
perform a uniform rotation of all the spins from the easy spin direction in
the crystal into a ``hard'' direction. The very small value of $\Delta $ in
the present system indicates that this is a ``soft'' ferromagnet, comparable
to very soft amorphous ferromagnets.\cite{amorphous} Finally we note that
the low temperature value of the spin stiffness constant $D(0)$ is $\sim
170meV-\AA ^2$. This gives a ratio of $D/kT_C\approx 7.9\AA ^2$ that is
quite large$,$ as might be expected for an itinerant electron system.\cite
{Ni}

As we raise the temperature towards $T_c$, for a conventional ferromagnet
that exhibits a second-order phase transition we expect the intensity of the
spin wave scattering at each $q$ to increase rapidly as the thermal
population increases according to the Bose factor. This thermal population
effect is further enhanced because of the renormalization (softening) of the
spin wave energy, with $D(T)$ expected to exhibit power law behavior and $%
D(T\rightarrow T_C)\rightarrow 0.$ The data in Fig. 2(a) show 
that the spin waves do soften with increasing temperature. However, they do
not collapse as $T\rightarrow T_C$, with $D(T_C)$ only about half of its low
temperature value. The measured (Bragg peak) magnetization as a function of
temperature is shown in Fig. 2(b) for comparison. This is rather typical of
a magnetization curve, and gives $T_C\approx 250K.$ However, the data are
not the same on warming and cooling, but exhibit an irreversibility of $5K.$
This irreversibility clearly demonstrates that the magnetic transition is
not a continuous, second-order transformation. We also found that in the
ordered phase the time scale for the magnetic intensity to equilibrate is
significant ($\sim 10-20$ min.), which provides further evidence that the
paramagnetic$\leftrightarrow $ferromagnetic transition is not continuous. At
lower $x$ we did not observe these long time scales, and the irreversibility
in the data was only $2K$ or less.

The spin dynamics associated with the phase transition for $x=\frac 13$ are
also anomalous as shown in Fig. 3, where we have plotted the magnetic
scattering at $q=0.07\AA ^{-1}$ for temperatures of $225$ and $235K$. These
data were taken with the same instrumental conditions as the data shown in
Fig. 1(a) and may be directly compared. At $225K$ we see that the spin waves
have softened somewhat in energy (shifting towards E=0), but the most
dramatic change is the appearance of a quasielastic component comparable in
intensity to the spin waves. At $235K$ the spin waves clearly renormalize to
lower energies (and broaden), but the central peak now dominates the
fluctuation spectrum. For typical isotropic ferromagnets such as Ni,\cite
{Nic} Co,\cite{Co} Fe,\cite{Fe} or amorphous materials\cite{amorphous} any
quasielastic scattering below $T_C$ is too weak and broad to be observed
directly in the data, and can only be distinguished by the use of polarized
neutron techniques.\cite{Polarized} This is also the case for smaller
concentrations of Ca where we have investigated the dynamics.

The data in Fig. 3 show that the observed width of the central peak is
broader than the spin waves, and this indicates that there is an intrinsic
width to the quasielastic scattering. The shape of the scattering can be fit
rather well by a Lorentzian lineshape for all three peaks, and Fig. 4(a)
shows the deconvoluted energy width $\Gamma $ for the quasielastic peak as a
function of $q$ at $T=240K$. The quadratic dependence on $q$ ($\Gamma
=\Lambda q^2$) demonstrates that this scattering originates from a diffusion
mechanism,\cite{HandH} with an effective diffusion constant $\Lambda =30$
meV-\AA $^2$ at this temperature. The $q$-dependence of the
energy-integrated intensity provides a measure of the associated length
scale for this diffusion, and from these data as well as from preliminary
small angle scattering (SANS) measurements we obtain a very short length
scale of $\sim 12\AA $ at this temperature. The SANS data also reveal that
as $T\rightarrow T_C$ from high $T$ the correlation length $\xi $ is only
weakly temperature dependent and does not diverge, with $\xi _{T_C}\sim
10\AA $. The short length scale and weak temperature dependence suggests
that this quasielastic scattering originates from electron diffusion on the
Mn sublattice.

We now turn to the temperature dependence of the intensities of scattering
as shown in Fig. 4(b). We see that the spin wave intensity does start to
increase with increasing temperature as expected, but above $\sim $200 K the
nature of the spin dynamics begins to change. In particular, as $%
T\rightarrow T_c$ the spectral weight for the spin wave excitations actually
starts to {\em decrease}, while the weight for the spin-diffusion component
rapidly increases. This spin diffusion component strongly dominates the
fluctuation spectrum near $T_c$, in marked contrast to conventional
ferromagnets. We have also found that the spectral weight for this spin
diffusion component can be shifted back into the spin waves by the
application of magnetic fields of a few tesla, which is the field regime
associated with the colossal magnetoresistance effects.\cite{Jin94} The
appearance of this quasielastic scattering for $x=\frac 13$ contrasts with
the behavior we find at x=0.15, 0.175, for example, where the magnetic
fluctuation spectrum below $T_C$ is dominated by spin wave excitations and
it is difficult to discern any distinct quasielastic component (as in
conventional ferromagnets).

It is therefore clear that this ferromagnetic phase transition in the $x=%
\frac 13$ material is not driven by the thermal excitation of spin waves as
for a conventional ferromagnetic phase transition, but rather is driven by
this spin diffusion mechanism. The short length scale associated with this
diffusion suggests that this component originates from the electronic
hopping from site-to-site on the Mn$^{3+}$/Mn$^{4+}$. The strong enhancement
of this component of the magnetic fluctuation spectrum for $x$ where the
colossal magnetoresistance effects are observed provides a direct link
between the ferromagnetic phase transition and the metal-insulator
transition associated with the electronic diffusion and localization that
occur in the same temperature/field regime. A straightforward interpretation
of the data is that the spin wave excitations and spin diffusion component
coexist, with a uniform magnetization throughout the sample. Another
interpretation of our results, though, is that the system is inhomogeneous,
consisting of two distinct phases. The phase preferred at low-T (or high-H)
is an ordered ferromagnet with metallic conductivity, a finite magnetization
and well-defined spin waves ($D(T_C)\sim \frac 12D(0)$),while the high-T
phase is a paramagnet where the electrons diffuse on a short length scale.
As the temperature is increased towards $T_C$ the fraction of these two
phases would then change as ferromagnetic phase converts to the paramagnetic
phase in a discontinuous fashion. We tend to favor this latter
interpretation since it is also consistent with the lattice anomalies and
magnetic irreversibilities we have observed for $x=\frac 13$. However, in
the small-$q$ regime explored in the present measurements we do not see any
direct evidence in the dynamics of the coupling of the magnetic system to
the lattice.\cite{Lattice}

We would like to thank S. M. Bhagat, L. Doloc, R. L. Greene, R. Ramesh, and
V. Venkatesan for helpful conversations. Research at the University of
Maryland is supported in part by NSF, DMR 93-02380.

Fig. 2. (a) Spin stiffness coefficient $D$ in $E=Dq^2$ as a function of
temperature. $D$ does not vanish at the ferromagnetic transition
temperature, in contrast to the behavior of a conventional ferromagnet.
Solid curve is a guide to the eye. (b) magnetization versus T obtained from
Bragg scattering. The data appear continuous and indicate $T_C\approx 250K$,
but the scattering is not reversible on warming and cooling.

Fig. 3. Inelastic spectrum at $225$ and $235K$ for $q=0.07\AA ^{-1}$. The
spin waves soften in energy with increasing temperature, but the dominant
effect is the development of the strong quasielastic scattering component in
the spectrum.

Fig. 4. (a) Width in energy $\Gamma $ versus $q$ for the quasielastic
scattering at $240K$. The quadratic dependence on $q$ indicates that this
scattering originates from spin diffusion. (b) Integrated intensities versus
temperature for the spin waves and spin diffusion scattering at $0.09\AA
^{-1}$. The spin wave intensity actually decreases as $T$ approaches $T_C$,
while the quasielastic scattering becomes an order of magnitude stronger
than the spin wave scattering. Above $T_C$ all the scattering in this range
of $q$ is quasielastic.


\begin{references}
\bibitem{anisotropy}  Indeed at doping concentrations less than x=$\frac 13$
we do observe some significant linewidths at low temperature.
\bibitem{Ni}  This is identical to Ni, for example. See ref. [11].
\end{references}
\end{document}